\documentclass[cits]{PoS}
\pdfoutput=1

\usepackage{verbatim}

\def\tr{\mbox{tr}}

\def\eq{Eqn.}

\def\fig{Fig.}

\newcommand{\ro}[1]{#1}
\newcommand{\bl}[1]{#1}

\title{Dressed Wilson loops as dual condensates
in response to magnetic fields}

\ShortTitle{Dressed Wilson loops as dual condensates
in response to magnetic fields}

\author{\speaker{Falk Bruckmann}, Gergely Endr\H{o}di\\
        Institut f\"ur Theoretische Physik, Universit\"at Regensburg, D-93040 Regensburg, Germany\\
        E-mail: \email{falk.bruckmann@physik.uni-regensburg.de}}

\abstract{We introduce dressed Wilson loops as a novel confinement observable. It consists of closed planar loops of arbitrary geometry but fixed area and its expectation values decay with the latter. The construction of dressed Wilson loops is based on chiral condensates in response to magnetic (and electric) fields, thus linking different physical concepts. We present results for generalized condensates and dressed Wilson loops on dynamical lattice configurations and confirm the agreement with conventional Wilson loops in the limit of large probe mass. We comment on the re\-nor\-ma\-li\-za\-tion of dressed Wilson loops.}

\FullConference{ The XXIX International Symposium on Lattice Field Theory - Lattice 2011\\
July 10-16, 2011\\
Squaw Valley, Lake Tahoe, California}

\begin{document}

We will present a construction \cite{Bruckmann:2011zx} in analogy to dressed Polyakov loops \cite{Bilgici:2008} (see also \cite{Gattringer:2006,Synatschke:2008}), where confining aspects of gauge theories are related to the response of the chiral condensate to an external magnetic field. Beside linking different physical effects, we expect this dressed observable to be free of additive renormalization, to depend nontrivially on the probe mass and to have applications beyond lattice QCD.

\begin{figure}[!b] 
\centering
  \hspace{-5cm}\includegraphics[angle=270,width=0.18\linewidth]{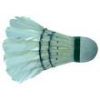}\hspace{-0.5cm}
  \includegraphics[angle=270,width=0.18\linewidth]{shuttlecock2.jpg}\hspace{-0.5cm}
  \includegraphics[angle=270,width=0.18\linewidth]{shuttlecock2.jpg}

  \includegraphics[width=0.3\linewidth,angle=90]{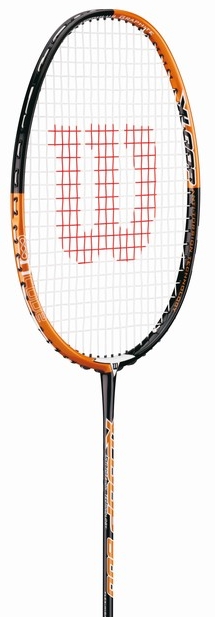}
\caption{A planar ``Wilson loop'' [Xloop 800 by 
Wilson Sporting Goods Co.] including a ``lattice'' is exposed to a constant external field perpendicular to it, the latter being symbolized by three ``arrows''.}
\label{fig racquet}
\end{figure}

Let us start by a geometric question, namely how to collect all planar Wilson loops (say on a lattice) with arbitrary geometry, but fixed area $\ro{S}$. Of course, the aim is to investigate the area law in such loops as a signal of confinement. The solution is rather simple, it is provided by an abelian and constant magnetic (or electric) field $\bl{b}$ perpendicular to the plane, cf.~Fig.~\ref{fig racquet}. The corresponding magnetic flux gives factors $e^{i\bl{b}\ro{S}}$ to all closed loops in that plane. Every gauge invariant quantity is made of such loops and so is the quark condensate. By a Fourier transform one arrives at a dual observable at fixed area.

To be more precise, we consider a constant abelian magnetic field along the $z$-direction, ge\-ne\-ra\-ted in the usual way by space-dependent gauge fields,
\begin{equation}
b_z=\partial_x a_y-\partial_y a_x\equiv \bl{b}\,,
\end{equation}
setting all electric charges to unity. All closed Wilson loops in the $(x,y)$-plane are modified as
\begin{equation}
W\,\to\,W\, e^{i\oint_C a}=W\, e^{i\int\!\!\int b_z}=W\, e^{\,i\bl{b}\ro{S}}\,,
\end{equation}
where we have made use of Stokes' theorem (and fixed an orientation). For space-like Wilson loops electric fields are used in the same manner\footnote{and therefore a general symbol $f$ was used for the external field in \protect\cite{Bruckmann:2011zx}}. 

Our considerations can be applied to any underlying space-time. On the lattice, the modifications of links to generate such an abelian field are known, see e.g.\ \cite{Al-Hashimi:2009}. Magnetic fields on a finite rectangle with extensions $L_x$ and $L_y$ are quantized in terms of the total area \cite{tHooft:1979uj},
\begin{equation}
\bl{b}=2\pi\,\frac{k}{L_x L_y}\,,\qquad k\in\mathbb{Z}\,,
\label{eqn quant}
\end{equation}
and on a discrete lattice with numbers of sites $N_{x,y}=L_{x,y}/a$ also bounded (related to the magnetic flux through a plaquette),
\begin{equation}
0 \leq k < N_x N_y\,,
\end{equation}
both in analogy to momenta on these spaces.

Let us denote the generalized quark condensate with links modified in this way by an index $\bl{b}$,
\begin{equation}
\Sigma_{\bl{b}}=\frac{1}{L_xL_y}\,\left\langle\tr\,\frac{1}{D_{\bl{b}}+m}\right\rangle\,.
\label{eqn sigma k}
\end{equation}
(where the trace includes an integral over $x$ and $y$ normalized by the area $L_xL_y$).
As a gauge invariant quantity this quark condensate consists of closed loops of all areas in the $(x,y)$-plane, such that schematically
\begin{equation}
\Sigma_{\bl{b}}=...\cdot 1+...\langle \mbox{plaquette}\rangle\cdot e^{\,i\bl{b}}
+...\langle W\big|_{\ro{S}=2}\rangle\cdot e^{\,2i\bl{b}}+...\,,
\end{equation}
where the dots stand for multiplicity factors, see also later.

The dual condensate after a (discrete) Fourier transform,
\begin{equation}
\tilde{\Sigma}(\ro{S}) \equiv \frac{1}{L_xL_y}\sum_{\bl{b}}
e^{-i\ro{S}\bl{b}}\,\Sigma_{\bl{b}}\,,
\label{eqn sigma s lattice}
\end{equation}
picks Wilson loops of fixed area\footnote{On a finite volume there is a ``contamination'' from loops closed through the boundary conditions, the effects of which can be suppressed \protect\cite{Bruckmann:2011zx}.} $\ro{S}$ and is therefore called `dressed Wilson loop'.

\begin{figure}[t!]
\centering
\includegraphics[width=0.64\linewidth]{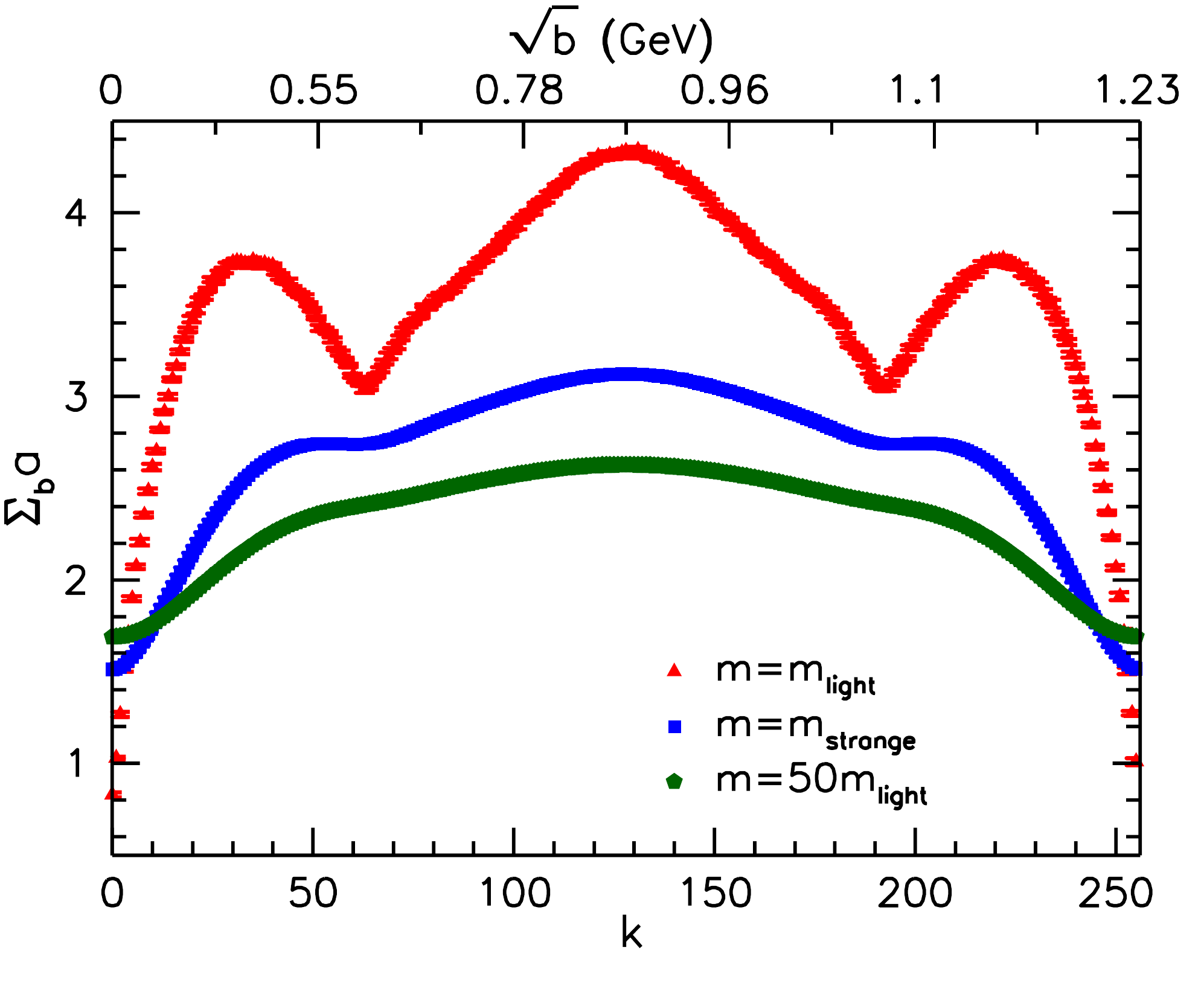} 
\caption{The generalized quark condensate $\Sigma_{\bl{b}}$, \eq~(\protect\ref{eqn sigma k}), in lattice units, in response to a constant magnetic field as a function of the flux quantum $k$ (see (\protect\ref{eqn quant})) for several probe masses. On the upper axis we also show the magnetic field in physical units. Dressed Wilson loops are the Fourier coefficients of these curves.}
\label{fig sigma k}
\end{figure}

At this point we would like to make several comments. First of all, the Dirac operator used in our construction is the two-dimensional one restricted to some $(x,y)$-plane (and is averaged over such planes finally). As a side effect its computation is numerically cheap. Secondly, we need for the Fourier transform all magnetic fields (not only those which might be of experimental relevance) as a tool on the level of observables, which means a partial quenching. Likewise the probe mass $m$ in the condensate can be different from the sea quark masses. Thirdly, other functions of the Dirac operator $D_{\bl{b}}$ different from the condensate (cf.~\cite{Synatschke:2008}) are possible to construct dressed Wilson loops, too.\\

Our numerical lattice results are based on $N_f=2+1$ stout smeared staggered\footnote{The staggered phases introduce particular signs in the dual condensates \protect\cite{Bruckmann:2011zx}, which for simplicity we avoid by always showing absolute values.} fermions with physical pion mass on $16^3\cdot 4$ lattices at $T=120$~MeV, for more details see \cite{Bruckmann:2011zx,Aoki:2006br}. We achieved a reasonable statistics from 5 configurations (needing 2.5 days on single CPU). 

In Fig.~\ref{fig sigma k} the generalized condensate $\Sigma_{\bl{b}}$ is shown as a function of the magnetic field $\bl{b}$. It grows with the magnetic field (`magnetic catalysis') and displays a ``shoulder'' structure, which is similar for the four-dimensional condensates (not shown). This behavior is quite different, when the magnetic field is included in the fermionic action \cite{D'Elia:2011,Bali:2011}. At half of the maximal magnetic flux the condensate saturates because of the periodicity due to discretisation. As expected the effects are washed out for heavy (probe) quarks.

\begin{figure}[b!]
\centering
\includegraphics[width=0.66\linewidth]{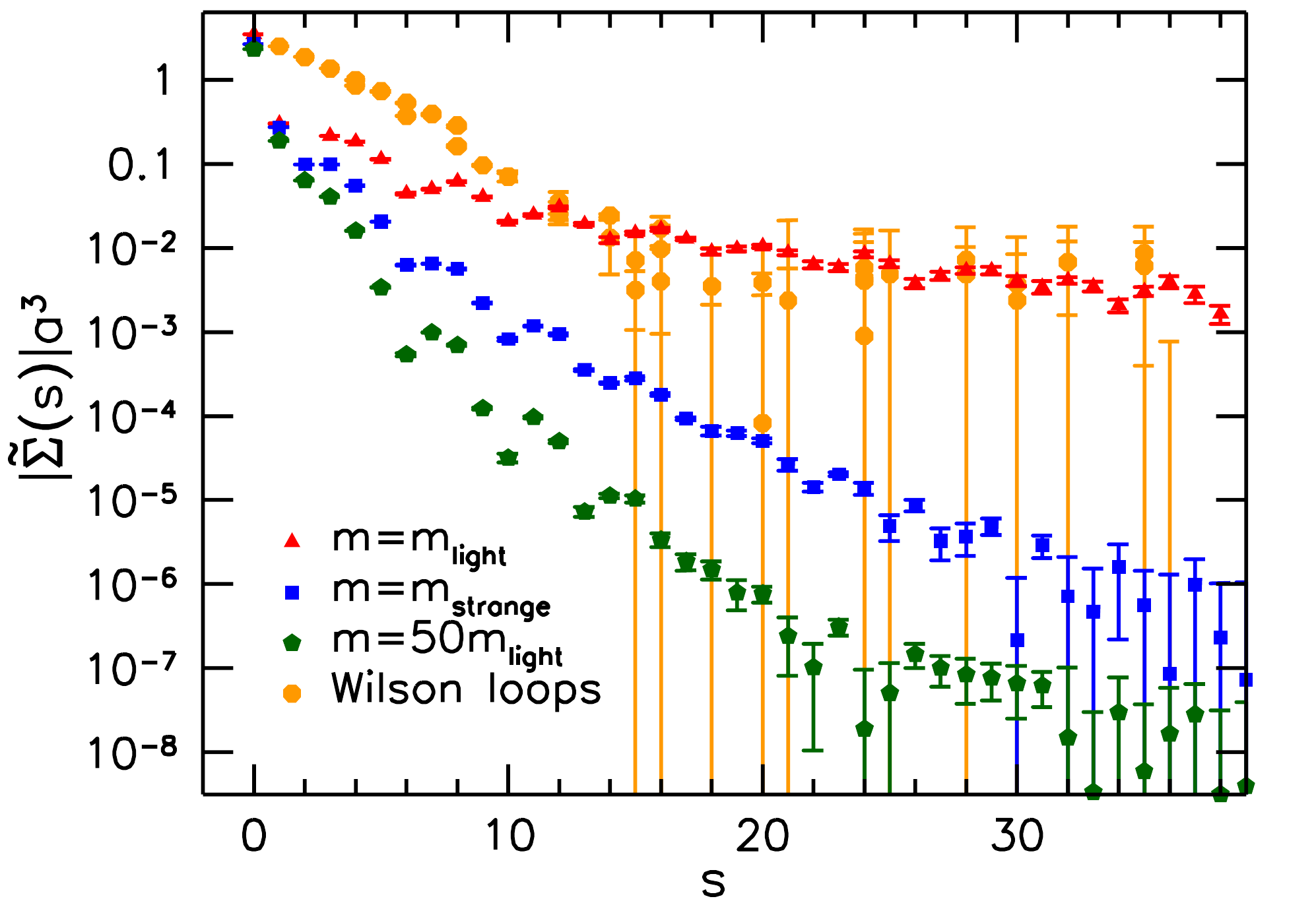} 
\caption{The dressed Wilson loops $\tilde{\Sigma}(\ro{S})$, \eq~(\protect\ref{eqn sigma s lattice}), in lattice units, dual to the condensates of \fig~\protect\ref{fig sigma k}, together with conventional Wilson loops  (the latter measured for different rectangles with side lengths up to $N_s/2$) as a function of the lattice area $s=S/a^2$.}
\label{fig sigma s}
\end{figure}

The Fourier transform of these data yields the dual condensates $\tilde{\Sigma}(\ro{S})$ depicted in Fig.~\ref{fig sigma s}. These dressed Wilson loops do decay with the area similar to conventional Wilson loops, however, with a different decay rate and some patterns in the dependence on the area $\ro{S}$. In order to clarify the situation, we now consider the geometry of lattice loops contained in this observable and make contact to conventional Wilson loops.

\begin{figure}[t]
\centering
\includegraphics[width=0.65\linewidth]{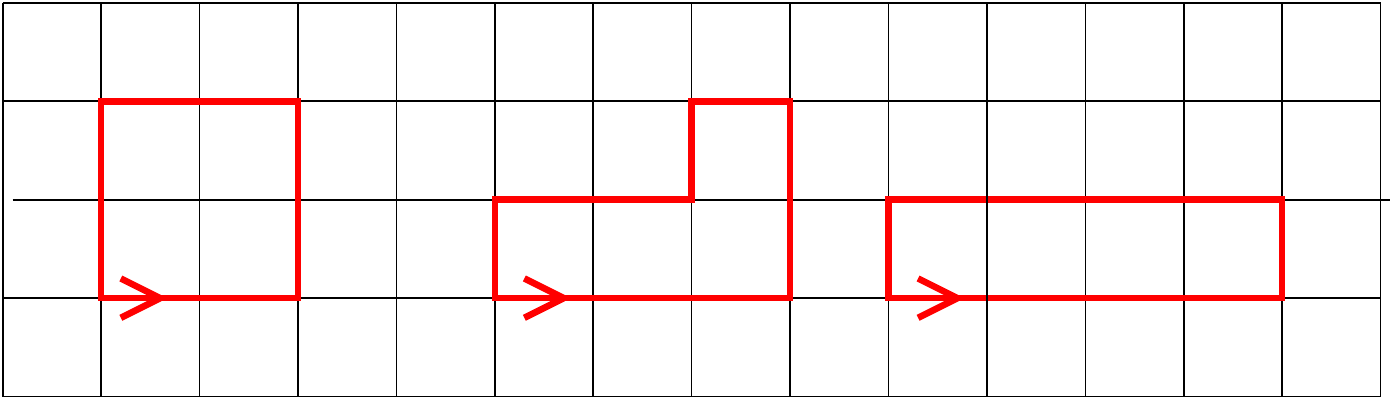}
\caption{Three lattice loops of area $\ro{S}=4a^2$ with different circumferences $l=8a,\,10a,\,10a$, respectively. The link products along these loops contribute to the dressed Wilson loop at this area, the latter two, however, with smaller weight factors at large probe masses, cf.~Eq.~(\protect\ref{eqn geom series}).}
\label{fig loops}
\end{figure}

The mechanism ensuring the latter is that long loops are suppressed by large probe masses (the same mechanism relates dressed and conventional Polyakov loops). In this limit one can expand the condensate in a geometric series of powers $l$ of the Dirac operator
\begin{equation}
m\:\tr\, \frac{1}{D_{\bl{b}}+m}=\tr\sum_{l}\frac{(D_{\bl{b}})^l}{m^{\,l}}
\label{eqn geom series}
\end{equation}
and for a lattice Dirac operator hopping over just nearest neighbors like the staggered operator we use, $l$ is just the circumference/length\footnote{Note that loops are not self-avoiding and can even go back and forth.} of the loop (in lattice units), see Fig.~\ref{fig loops} for examples. Obviously, long loops are suppressed by the corresponding factor of inverse mass. 

The `ideal lattice loops' in the sense of maximizing the area at fixed circumference can -- by virtue of the following loop modifications

\begin{figure}[!h]
\centering
\includegraphics[height=0.12\linewidth]{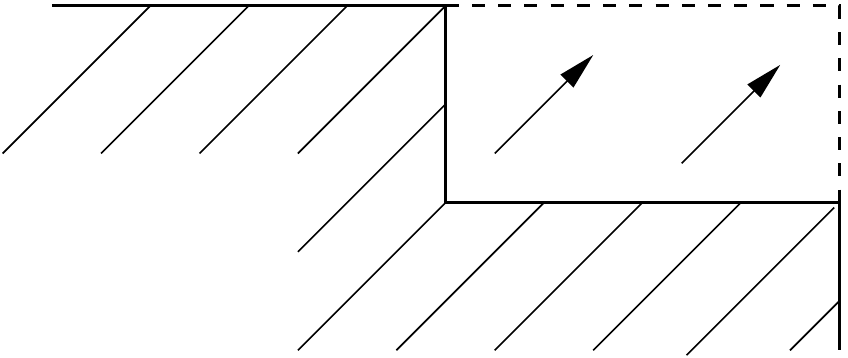}\qquad
\includegraphics[height=0.12\linewidth]{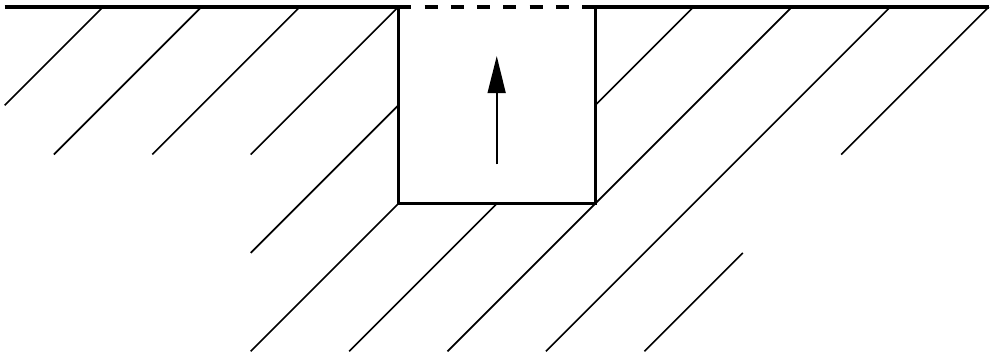}
\end{figure}
\noindent -- easily be shown to be rectangles (in contrast to circles in the continuum).

The `entropy', i.e.\ the multiplicity of different loops in the observable, may counteract the suppression away from the ideal loop. It actually does not do so, since first of all there are no disconnected loops in $\tr\, D^l \propto D(x,y)D(y,z)\ldots D(w,x)$ and secondly we have shown that the growth of the multiplicity factors (listed in the appendix of \cite{Bruckmann:2011zx}) at fixed area as a function of the circumference is always exceeded by the mass suppression. We conclude that for large mass the dual condensate $\tilde{\Sigma}(\ro{S})$ contains predominantly loops of small circumference, whose shapes are (close to) rectangular. 

\begin{figure}[!t]
\centering
\includegraphics[width=0.66\linewidth]{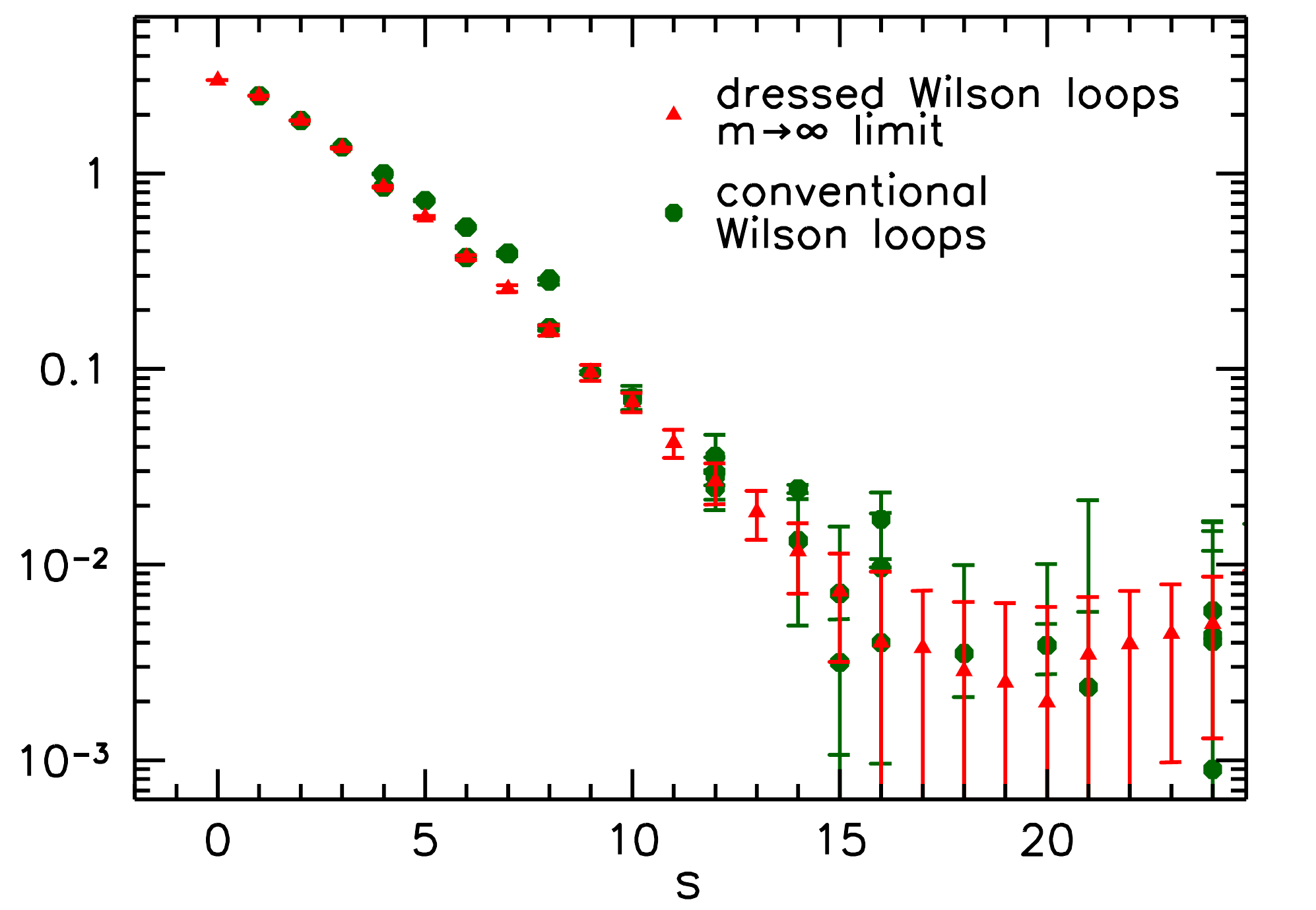}
\caption{The leading term in the large mass limit of the dressed Wilson loop, in comparison to conventional Wilson loops.}
\label{fig sigma s largemass}
\end{figure}

In the large mass limit we can identify the leading contribution to the dressed Wilson loop at area $\ro{S}$, which is the quantity dual to the minimal power $l_{{\rm min}}(\ro{S})$ of the Dirac operator, and divide out the multiplicity and mass factors, for details see \cite{Bruckmann:2011zx}. The result is plotted in Fig.~\ref{fig sigma s largemass} showing agreement with the conventional Wilson loops within error bars.\\

An intuitive physical picture of our construction can be drawn from Wilson loops extending in space and time, which represent a quark-antiquark pair; the corresponding dressed Wilson loops are obtained from an external electric field. In the common picture of confinement the quark and the antiquark are bound together by nonabelian forces, that result in a potential  growing linearly with the distance. Our (Euclidean) electric field acts on the quark and antiquark, as they carry opposite electric charges, and thus indeed attempts to vary the distance between them. Therefore it is clear that the electric field probes the nonabelian forces. Our formalism says that the latter are exhibited from the response of QCD quantities to all such electric fields (similar to inverse scattering).

We want to conclude with two features of dressed Wilson loops concerning the IR and UV behavior, that are again very similar to those of dressed Polaykov loops \cite{Bruckmann:2007b,Bilgici:2008,Synatschke:2008,Bruckmann:2008sy,Bilgici:2010,Zhang:2010}. Both rely on the `fuzziness', which makes this observable less sensitive to the lattice spacing. 

In a spectral representation of a Dirac operator with purely imaginary eigenvalues\footnote{neglecting nongeneric zero modes},
\begin{equation}
\Sigma_{\bl{b}}=\frac{1}{V}\,\left\langle\tr\,\frac{1}{D_{\bl{b}}+m}\right\rangle
 =\frac{1}{V}\,\left\langle \sum_{\lambda_{\bl{b},i}>0}\frac{2m}{\lambda_{\bl{b},i}^2+m^2}\right\rangle
\end{equation}
it is obvious that the dominant contribution to all condensates comes from the IR part $\lambda\lesssim m$, which is lost for conventional Wilson loops as $m\to\infty$.

The renormalization properties of the generalized condensate we conjecture to be the same as those of four-dimensional condensates. Since their divergences do not depend on the external field, they are removed by the Fourier transform and therefore additive renormalization is not necessary in the dual condensates (for non-vanishing area). The multiplicative renormalization can easily be accounted for by multiplying the condensate by the bare mass $m$. How this translates in the large mass limit into the renormalization of conventional Wilson loops and how dressed Wilson loops at finite mass can be interpreted and further explored as a tool, are interesting aspects of our construction to be investigated further.\\

The authors thank the Budapest-Wuppertal collaboration for the permission to use their staggered code for the configuration production. This work has been supported by DFG (BR 2872/4-2) and the Research Executive Agency 
of the European Union 
(ITN STRONGnet).

\end{document}